\documentclass{aastex}
\usepackage{emulateapj5,graphicx}
\begin{document}

\newcommand{\bm}[1]{\mbox{\boldmath $#1$}}
\title{
Regulated star formation in forming disk galaxies\\
 under ultraviolet radiation background
}
\author{Hajime Susa\altaffilmark{1}
\vskip 0.2cm
\affil{Department of Physics, Konan University, Okamoto, Kobe, Japan}
\altaffiltext{1}{susa@konan.ac.jp}
}
\begin{abstract}
We perform radiation hydrodynamics simulations on the evolution of
 galactic gas disks irradiated by ultraviolet radiation background.
We find gas disks 
with $N_{\rm H} \ga 10^{21} {\rm cm^{-2}}$ exposed to
 ultraviolet radiation at a level of $I_{\rm 21}=1$ can be self-shielded
 from photoheating, whereas the disk with  $N_{\rm H}
 \la 10^{21} {\rm cm^{-2}}$ cannot. We also find that the unshielded
 disks keep smooth density distribution without any sign of
 fragmentation, while the self-shielded disks easily fragment into small
 pieces by self-gravity, possibly followed by star formation. 
The suppression of star formation in unshielded disks
is different from
photoevaporation effect, 
since the assumed dark halo potential is deep enough to keep the
 photoheated gas. 
Presence of such critical threshold column density would be one of the
 reason for the so-called down-sizing feature of present-day galaxies.
\end{abstract}


\section{Introduction}
\label{intro}

Recent advances of available computational resources enable us to
simulate the entire disk galaxy to investigate parsec scale structures in the
disk\citep{Wada07,Robertson07,Tasker08,Saitoh08}. 
In addition, numerical techniques on Radiation HydroDynamics (RHD)
simulations are now being developed, especially in the field related to
the cosmic
reionization\citep{GA01,C2ray,Saru,Susa06,TSU3,Yoshida07,Qiu07,Whalen08,Altay08}. 
Using these schemes, we are now ready to tackle the
numerical simulations of the detailed structures in galactic disks 
coupled with radiation transfer.

The importance of ultraviolet radiation transfer effects on the galaxy
formation/star formation in galaxies
have been pointed out by several authors.
First of all, 
the observed star formation threshold in galactic disks \citep{K89,MK01}
could be explained by the self-shielding effects of galactic disks from
external ultraviolet radiation field\citep{Schaye01,Schaye04}.
\citet{Schaye04} obtained the physical state of the disks exposed to
external ultraviolet field, using one dimensional calculation with CLOUDY\citep{CLOUDY}.
Based upon the liner stability arguments\citep{Toomre}, 
he has demonstrated that the inner parts of the galactic disks shielded
against external radiation field by dust absorption could be
gravitationally unstable, whereas the unshielded outer parts of the disks
are stable. These results nicely explain the observed features of
galactic disks, however, it has not been tested 
by multi-dimensional RHD simulations.

Secondly, in the context of galaxy formation, radiative feedback by
ultraviolet background is expected to be very important especially for
low mass galaxies. The photoheating can heat the gas up to
$10^4$K, which prevent the gas from collapsing, in case the gravitational
potential of the dark matter halo is not deep enough to retain the
photoheated gas. Such feedback
effects are quoted as photoevaporation, which is well studied at various
levels\citep{UI84,Efs92, BR92,TW96,BL99,FT00,Gne00c,Kita00,Kita01,SU04a,SU04b}.
In addition, there are some evidences in dwarf galaxies that the
star formation is suppressed in ``hot'' ($\sim 10^4$K) phase \citep{YL97a,YL97b}, 
which infer that photoheating can suppress the
star formation in galaxies, even if the potential of
the dark halo is deep enough to prevent the gas from evaporating.
This issue could have a concern to the so-called down-sizing
problem in nearby galaxies\citep{Cowie96,downsizing}. They found
that old galaxies are massive, while the young ones are less massive.
Such trend continues to higher
redshift\citep[e.g.,][]{downsizing_subaru,Erb06,Reddy06,Papovich06}.
One of the possible interpretation of this feature is 
that star formation proceed very rapidly in
massive galaxies, whereas it is a slow process in less
massive galaxies for some reason. 
\citet{downsizing} also pointed out that these two groups are well
defined, i.e. they have a clear boundary at $\sim 10^{10}M_\odot$ in stellar mass. This critical mass scale is
too large to be related to the photoevaporation mechanism, however,
photoheating could still be the candidate to explain the down-sizing
mass if we can show the mechanism to suppress the star formation even
in halos in which the photoionized gas can be inherent.

In this paper, we examine the fragmentation of gas disks embedded in dark
halo potential under the ultraviolet radiation field, using
the recently developed RHD code RSPH\citep{Susa06}. This paper is organized as
follows. In the next section, the numerical scheme is briefly
summarized. In section \ref{setup}, the setup of numerical simulations
are described. We show the results of numerical simulations as well as
the analytic estimate in section \ref{results}.
Sections \ref{discussion} and \ref{conclusion} are
devoted to discussions and conclusion. 

\section{Methodology}
\label{scheme}
We perform numerical simulations by Radiation-SPH(RSPH) code \citep{Susa06}.
 The code can compute the fraction of primordial chemical species e$^-$, H$^+$, H,
H$^-$, H$_2$,  and H$_2^+$ by fully implicit time integration. It also
can deal with multiple sources of ionizing radiation, as well as
the radiation at Lyman-Werner band.

Hydrodynamics is calculated by Smoothed Particle Hydrodynamics (SPH)
method. We use the version of SPH by \citet{Ume93} with the modification
on the SPH kernel function and symmetrization of equation of motions 
according to \citet{SM93}. We also adopt the particle resizing
formalism by \citet{Thac00} in which the number of neighbor SPH
particles are kept almost constant without sudden changes.

The non-equilibrium chemistry and radiative cooling 
for primordial gas are calculated by the code
developed by \citet{SuKi00}, where H$_2$ cooling and
reaction rates are mostly taken from \citet{GP98}.
As for the photoionization process, 
we employ so-called on-the-spot approximation
\citep{Spitzer78}.
In this paper, we also added the radiative cooling due to metals employing the
formula given in \citet{DM73}, assuming $Z=10^{-2}Z_\odot$.
The metallicity of nearby disk galaxies are normally larger than
$Z=10^{-2}Z_\odot$, 
however, the observed metallicity of the IGM is $Z
\sim 10^{-2}Z_\odot-10^{-3}Z_\odot$ \citep[e.g.,][]{Son01} 
which is the ingredient of the galaxies. Since we are interested in the
forming disk galaxies, we employ $Z=10^{-2}Z_\odot$ in the present set
of simulations.
Remark that the radiative transitions of heavy elements
are not the dominant process of radiative cooling for $\ga
1000$K \citep{BH89,SU04a} in case $Z=10^{-2}Z_\odot$ is assumed. 
H and H$_2$ molecules are the main coolant of such
gas\footnote{We have to keep in mind that for low temperature ($T\la
500$K), or high density ($n_{\rm H}\ga 10^4{\rm cm^{-3}}$) realm, even
$Z\sim 10^{-4}Z_\odot$ could be important to decide the fate of cooling gas\citep{Omukai00}. }.

The optical depth is integrated utilizing the 
neighbor lists of SPH particles.
In our old scheme \citep{SU04a}, we create many grid points on the light
ray between the radiation source to an SPH particle.    
In the present scheme, we do not create so many grids. 
Instead, we create one grid point on the light ray per one SPH
particle on its upwind. 
We find the neighboring 'upstream' particle for each SPH particle on its
line of sight to the source, which corresponds to the grid point.
Then the optical depth from the source to
the SPH particle is obtained by summing up the optical depth at the 'upstream'
particle and the differential optical depth between the two
particles. The more details are described in \citet{Susa06}.
We assess the optical depth for ionizing photons as well as the
Lyman-Werner band photons by the method described above.
In the present version of the code, the dust opacity is not included in
the calculation.

The code is tested for various standard RHD problems\citep{Susa06}, and
already applied to the issues on the radiative feedback effects of first
generation stars \citep{SU06,Susa07}. 
We also take part in the code comparison
project with other radiation hydrodynamics codes \citep{TSU3} in
which we find reasonable agreements with each other.

\section{Setup of Numerical simulations}
\label{setup}

We perform numerical simulations of galactic gas disk embedded in a
dark halo potential. The dark halo potential is fixed as 
\begin{equation}
\Phi_{\rm DH}(r) \equiv
 -\left(\frac{27}{4}\right)^{1/2}
\left[
\frac{av_1^2}{\left(r^2+a_1^2\right)^{1/2}} +
\frac{av_2^2}{\left(r^2+a_2^2\right)^{1/2}} 
\right],
\label{eq:DHpot}
\end{equation}
where $r$ denotes the distance form the center, 
$a=1{\rm kpc},a_1=0.3{\rm kpc},a_2=5{\rm kpc}$ and $v_1=v_2=100{\rm km/s}$. 
Thus the assumed potential is similar to the one employed by
\citet{Wada07}, except that the rotation velocity at given radius 
is smaller by a factor
of 2,
since we are interested in the forming galaxies relatively
less massive than our Galaxy.

In the present set of simulations, we assume initially uniform disk
(Fig. \ref{fig:model}) 
with slight perturbations (displacements of SPH particles are $\la$ 10 \%). 
The initial thickness and radius of the disks, $H_{\rm i}$, $ R_{\rm
disk}$ are 100pc and 3kpc for all of the runs, respectively.
The initial velocity of the gas particles in the disk are assumed to be the 
Kepler rotation velocity around the dark halo potential center.
Using equation (\ref{eq:DHpot}), the rotation velocity at the edge
of the disk $r=R_{\rm disk}$ is as large as  $\sim 100$km/s. Thus, the
dark halo potential is deep enough to retain the gas even if it is
photoheated to $\sim 10^4$K.
The initial densities of the disks, $\rho_{\rm i}$, are changed  depending
on the models, those are listed in Table \ref{tab:models}.

The light rays of external ultraviolet radiation field are assumed to be
perpendicular to the gas disk. We consider two
directions (upward/downward) as shown in Fig. \ref{fig:model}. 
This particular choice of the ray-direction is an approximation
      if we consider background radiation field exactly, since background
      photons are coming from all angles. If we take into account
      such effects, the boundary between the shielded and unshielded
      regions are softened, which might change the self-shielding effects
      slightly. However, as far as we keep the number of photons coming
      into the disk unchanged, the self-shielding condition is also
      almost unchanged, since it is determined by the balance between
      the number of recombining hydrogen atoms and the number of photons.
The
flux of the field at the Lyman limit is given so that the mean intensity
at the midplane of the disk equals to $I_{\rm 21}\times 10^{-21} {\rm erg
s^{-1}cm^{-2}Hz^{-1}str^{-1}}$ in case the opacity of gas disk is ignored.
We consider an AGN type spectrum, which is proportional to $\nu^{-1}$. 

For all of the runs, we employ common physical/numerical parameters
of gas disk listed in Table\ref{tab:common}, while the model dependent
parameters are listed in Table \ref{tab:models}. As shown in the table,
we perform ten runs changing the density of the disk and the intensity
of the radiation field. 

Since the resolution of the simulation is constrained by the mass of an SPH
particle, we have the critical density above which the gravitational 
fragmentation
are not captured properly. According to \citet{BB97}, the density $n_{\rm H,res}$ is given as
\begin{equation}
n_{\rm H,res}=\frac{3}{4\pi m_{\rm p}}\left(\frac{5k_{\rm B}T_{\rm min}}{2 G
 m_{\rm p}}\right)^3\frac{1}{\left(2 m_{\rm sph} N_{\rm nei}\right)^2}\label{eq:den_res}
\end{equation}
where $m_{\rm sph}, N_{\rm nei}, T_{\rm min}$ denote the mass of each
SPH particle, number of neighbor particles and the minimal temperature
set in the simulations, respectively.
The symbols $k_{\rm B},G,m_{\rm p}$ have ordinary meanings, i.e., they represent
Boltzmann constant, gravitational constant and proton mass, respectively.
In the present set of simulations, $n_{\rm H,res}$ equals to  $235{\rm
~cm^{-3}}$, except the runs B/2,Br/2,B/8 and Br/8 in which the mass of a SPH
particles is larger than other regular runs 
by a factor of two or eight (see Table \ref{tab:models}).
The softening length of
gravitational interaction between SPH particles , $\epsilon$, is set so
as to satisfy
\begin{equation}
\epsilon=\frac{1}{2}\left(\frac{3N_{\rm nei}m_{\rm SPH}}{4\pi m_{\rm p}n_{\rm H, res}}\right)^{1/3}.\label{eq:softening}
\end{equation}
This expression guarantees that the self-gravity of a dense clump above $n_{\rm H, res}$ is softened.

Given these configurations, we can assess the Toomre's $Q$ value to
understand the stability of the initially uniform disk.
 Toomre's $Q$ is defined as 
\begin{equation}
Q=\frac{c_{\rm s}\kappa}{\pi G \Sigma}.
\end{equation}
Here $c_{\rm s},\kappa$ and $\Sigma$ denote the sound speed of the gas,
epicyclic frequency and the surface density of the disk. In case $Q>1$
holds, the disk is gravitationally stable because of the Coriolis
force/thermal pressure, 
whereas it is unstable if $Q<1$ is satisfied\citep{Toomre}.

The actual
$Q$ parameters for three models A, B and C (see Table \ref{tab:models}) are plotted in Fig.\ref{fig:Q}. Here
$c_s$ is assumed to be the sound speed corresponding to $T_{\rm min}=300$K.
Since we assume uniform disks for all models, the disks are more
unstable at the outer radii. In the present simulations, the radius of the disks are 3kpc for all models. Thus, there exist critical radii
above which the disks are unstable if those are cooled down to $T_{\rm
min}$, 
although the unstable region is narrow for model A ( it is unstable only
at the edge of the uniform disk). 
It is worth noting that if we employ $T_{\rm min}$ lower than $300$K, the
unstable region should expand, since the gas temperature will go down to
$T_{\rm min}$, that results in smaller $Q$.

\section{Results}
\label{results}
\subsection{Stability of the disks without Radiation}
First, we present the results for $I_{21}=0$,
i.e. no external radiation field. The face-on/edge-on view of the
snapshots for the disks in models A, B and C are shown in
Fig.\ref{fig:montage_norad}. 
The snapshots are taken at $t=300$Myr for model A, $t=120$Myr
for model B and $t=40$Myr for model C.
The particular choices of these times when the snapshots are taken
basically corresponds to the time when the time steps at dense gas clumps
collapsed below $n_{\rm H,res}$ becomes so short ( $\la 10^3$yr ) that
physical time in numerical computation evolve very slowly. Since the
growth time scale of gravitational instability is shorter in denser
disks, we take the earlier snapshot in model C than others.

It is clear that the disks in all models fragment into small
filaments/knots. At the same time, the inner parts of the disks are
stable. As shown in Fig.\ref{fig:Q}, $Q$ values are
larger than unity at smaller radii, whereas they are smaller than unity at
larger radii. Therefore, the inner parts of the disks are stable, which is consistent
with the present results. In addition, the critical radii outside which
the disks are unstable predicted by Toomre criterion are also
consistent with numerical results. Indeed, the critical radii read from
Fig.\ref{fig:Q} are 2.8kpc, 1.6kpc and 0.8kpc for models A,B and C, those
are almost consistent with the boundary radii of smooth parts of the
disks.

In Fig.\ref{fig:phase_norad}, the phase diagram ($n_{\rm H} - T$ plane) of the three runs at the
same snapshots as in Fig.\ref{fig:montage_norad} are shown. 
The gas temperature in
all of the models are cooled efficiently, 
which are almost close to $T_{\rm min}=300$K at high
density realm ($n_{\rm H} \ga 10^3 {\rm cm^{-3}}$). Such efficient
cooling justify the stability argument based upon the $Q$ values
plotted in Fig.\ref{fig:Q} in which $T=T_{\rm min}$ is assumed.

These phase digram also show that very dense regions above $n_{\rm H,
res}$ are formed in all of the models. These dense regions correspond
to the fragments presented in Fig. \ref{fig:montage_norad}. 

\subsection{Effects of Ultraviolet Radiation on the Disk Stability}
Now, we show the snapshots of the runs with $I_{21}=1$ (Ar, Br and Cr)
in Fig.\ref{fig:montage_rad}, i.e. $t=300$Myr for model Ar, $t=120$Myr
for model Br  and $t=40$Myr for model Cr.
The snapshots are synchronized to the corresponding ones in
Fig.\ref{fig:montage_norad}.
At a first glance we find clear difference from the runs without
ultraviolet radiation. In model Cr, we find small filaments and knots
as were found in model C, whereas we cannot find any sign of
fragmentation in models Ar and Br. In addition, the density of the disks
in models Ar and Br becomes lower than that in models A and B. We also
observe that the disks are geometrically thicker than the previous ones.
In Fig.\ref{fig:phase_rad}, the phase diagrams are shown for models Ar,
Br and Cr. In all of the runs, significant amount of materials are
photoheated up to $\ga 10^4$K in low density region, that contribute to
the disk thickening. Additionally, 
no very dense gas component above $n_{\rm H,res}$ is found in models Ar
and Br, while we find it again in model Cr as was found in model C. 
Thus, these results indicate that the presence of external ultraviolet radiation field suppress the
fragmentation of the disks with low density models (Ar and Br), although
it does not for high density model (Cr).

\subsection{Self-Shielding from Ultraviolet Radiation}
The difference between low density models(Ar, Br)
and high density model(Cr) comes from the self-shielding
effects. Fig.\ref{fig:z-X}  shows the density (top), temperature
(middle) and HI fraction $y_{\rm HI}$(bottom) distributions along the direction
perpendicular to the disks. In model Cr (right column), the gas near the
midplane ($z=0$) is neutral, cold and dense, because of the self-shielding.
On the contrary, in models Ar and Br, 
cold and dense regions near the midplane is
relatively smaller than that in model Cr. To be more quantitative,  
Fig.\ref{fig:pdf_tmp} shows the temperature probability
distribution functions (PDFs)for models Ar,Br,Cr (top) and A,B,C(bottom). 
The temperature PDFs show the mass fraction of the disk
found within a logarithmic temperature bin $\Delta(\log T)=0.06$.
In the runs without radiation field (A,B and C), most of the mass is
condensed in the coldest phase at $T_{\rm min}$.
On the other hand, 
the temperature PDFs are broader in the runs with radiation.
In models Ar and Br, most of the mass in the disks exist in the hot
phase ($\sim 10^4$ K), while cold phase ($< 10^3$K) is dominant in
model Cr.
Such large difference in temperature distribution results in different stability of
the disks.
According to Fig. \ref{fig:Q}, $Q$ values for models Ar and Br satisfy
$\ga 0.5$ if $r\le 3$kpc and $T=T_{\rm min}$. In actual
simulations, we find $T\sim 10^4$K, which means that $Q$ values are larger by a factor
of $\sqrt{10^4/300}=5.77$ than 0.5, i.e. $Q > 1$ is achieved everywhere
in the disks for these models. Thus,
these disks are stable, while the disk in model Cr is unstable because
of its coldness.

A rough estimate of 
self-shielding criterion for primordial gas has been derived by
\cite{SU00a}, who evaluate the shielded photoheating rate to
compare with the peak H$_2$ cooling rate below $10^{4}$K. Remark that
 heavy elements are not the dominant coolant for $10^3 {\rm K} \la T\la 
 10^4{\rm K}$, in case $Z\la 10^{-2}Z_\odot$\citep{SU04a}.

Similar
argument is also found in \cite{Corbelli}, which describes the thermal
instability of the primordial gas.
According to \citet{SU00a}, the photoheating rate per unit volume 
at the midplane of the disk for
the optically thick limit is given as $n_{\rm H,c}y_{\rm
HI,c}\mathcal{H}$  where 
\footnote{The expression $\mathcal{H}$ is 2
times larger than the equation (A12) in \citet{SU00a}, since the
radiation field irradiate the right side of the disk as well as the
reverse side in the present simulations.}
\begin{equation}
\mathcal{H}=\frac{4\pi I_{\nu_{\rm L}}\nu_{\rm
 L}\sigma_{\nu_{\rm
 L}}}{3}\frac{\Gamma\left(\beta\right)}{1+\beta}\tau_{\nu_{\rm L}}^{-\beta}
\label{eq:photoheating}
\end{equation}
Here $n_{\rm H, c}$ and $y_{\rm HI,c}$ are the
number density of the hydrogen nucleus and HI fraction at the midplane
of the disk, $\sigma_{\nu_{\rm L}}$ is the photoionization cross section
at Lyman
limit, $\nu_{\rm L}$ is the Lyman limit frequency, $I_{\nu_{\rm L}}$
denotes the incident ultraviolet intensity at Lyman limit,  $\Gamma(x)$ 
is the gamma function. $\beta$ used in this
equation is defined as $\beta\equiv 1+\left(\alpha-1\right)/3$,
where
$\alpha$ denotes the spectral index, i.e., $I_{\nu}\propto \nu^{-\alpha}$
is assumed. $\tau_{\nu_{\rm L}}$ denotes the optical depth at Lyman
limit, which is written as,
\begin{equation}
\tau_{\nu_{\rm L}} = \frac{\left\langle y_{\rm HI}\right\rangle N_{\rm
 H} \sigma_{\nu_{\rm L}}}{2}.
\label{eq:tau}
\end{equation}
Here $N_{\rm H}$ denotes the column density of the disk, $\left\langle
y_{\rm HI}\right\rangle$ is the HI fraction averaged along the direction
perpendicular to the disk.
The peak H$_2$ cooling rate per unit volume below $10^4$K is described as 
$n_{\rm H,c}^2\mathcal{C}_{\rm H_2}$, where
$\mathcal{C}_{\rm H_2}\simeq 10^{-26}{\rm erg~s^{-1}~cm^{3}}$\citep{SK87,SU04a}.
In addition, we assume hydrostatic equilibrium in the vertical direction. 
As a result, the number density of hydrogen nucleus at the midplane of
the disk is related to the column density as follows:
\begin{equation}
n_{\rm H,c} = \frac{\pi G N_{\rm H}^2 \mu m_{\rm p}}{2 c_{\rm s}^2}.
\label{eq:zbalance}
\end{equation}
Here the gas disk is assumed to be isothermal with sound velocity
$c_{\rm s}$.
Letting the photoheating rate be equal to the
H$_2$ peak cooling rate, combined with the equations (\ref{eq:photoheating}), (\ref{eq:tau})
and (\ref{eq:zbalance}), we obtain the threshold column density $N_{\rm
H,sh}$, above which the photoheating is shielded enough for the midplane
to cool down to $\la 1000$K:
\begin{equation}
N_{\rm H,sh}=\left(\frac{2^{\beta+3} c_{\rm s}^2 I_{\nu_{\rm L}}\nu_{\rm
 L}\sigma_{\nu_{\rm L}}^{1-\beta}\Gamma\left(\beta\right) }{3G \mu
 m_{\rm p}\mathcal{C_{\rm
 H_2}}y_{\rm HI,c}\left\langle y_{\rm HI}\right\rangle^\beta\left(\beta+1\right)}\right)^{1/\left(2+\beta\right)}
\end{equation}
Using the present assumptions ($\alpha=1$, $I_{21} = 1$), we have
\begin{eqnarray}
N_{\rm H,sh}\simeq 2.5 &\times& 10^{21} {\rm cm^{-2}} \nonumber \\
 &\times &\left(\frac{I_{21}}{1}\right)^{1/3}\left(\frac{c_{\rm s}}{10{\rm
  km s^{-1}}}\right)^{2/3}\left(\frac{\left\langle y_{\rm
			   HI}\right\rangle }{0.5}\right)^{-1/3}
\label{eq:Nsh}
\end{eqnarray}
Here we also assume $y_{\rm HI,c}=1$ in the above expression, since it
is true for all models Ar, Br and Cr.

On the other hand, the initial column density of the disks in models
Ar, Br and Cr are $6.3\times 10^{20}{\rm cm^{-2}}$, $1.3\times
10^{21}{\rm cm^{-2}}$ and $3.8\times 10^{21}{\rm cm^{-2}}$,
respectively. Thus, the numerical simulations indicate $1.3\times
10^{21}{\rm cm^{-2}} \la N_{\rm H,sh} \la 3.8\times 10^{21}{\rm
cm^{-2}}$, which almost agrees with the rough analytic estimate shown above.

\subsection{Probability Distribution Functions of Density Field}
The differences among the six models (A,Ar,B,Br,C and Cr) are clarified
in Fig. \ref{fig:pdf} in terms of density probability distribution
functions(PDFs). Six panels represent the snapshots of the density PDFs
for six models. Here the density PDFs are defined as the mass fraction
found within a logarithmic density bin $\Delta(\log n_{\rm H})=0.18$.
The histograms drawn by thin lines represent the initial PDFs, while
the thick lines show the evolved ones.

It is clear that in the models A,B,C, and Cr,  
the PDFs extend beyond the resolution limit $n_{\rm H, res}$. Thus, dense
self-gravitating fragments formed in these disks, which would lead
to the star formation activities.

It looks strange at first that very dense clumps form beyond $n_{\rm H,
res}$, although the gravitational force is softened for $n_{\rm H} \ga n_{\rm H,
res}$ (see eq. (\ref{eq:softening})). 
In fact, even if gravitational force could be
neglected at $n_{\rm H} \ga n_{\rm H, res}$, matters accrete from the
outer envelope.
As a result, density (or
pressure) of the clump becomes higher than $n_{\rm H, res}$, in order to
compensate for the increasing ram/thermal pressure of accreting material.

On the other hand, in the models Ar and Br, PDFs have very sharp cut-off below
$n_{\rm H, res}$ for the snapshot at $t=300$Myr(Ar) and $t=120$Myr(Br),
although models  A and B have
clear fragmentation signature at the same snapshots. 
Thus, the impression from the
montage of the disk is correct, i.e. the disk in models Ar and Br 
do not fragment into dense clouds. This result could be interpreted
that the star formation in these disks are heavily suppressed by ultraviolet
radiation field.

\subsection{Convergence of PDF}
The convergence of density PDF is checked for models B and
Br. We perform runs with larger SPH particles mass by a factor of two
and a factor of eight.
These runs are
tagged 
as B/2, Br/2 and B/8, Br/8
(Table \ref{tab:models}). Fig.\ref{fig:conv} shows the density PDFs for models
B,B/2,B/8 (top) and models Br,Br/2,Br/8 (bottom) at
$t=40,100,120$Myr. In the half resolution runs (B/2 and Br/2), the density
resolution limits are four times smaller than those in B and Br, because
$n_{\rm H, res}$ inversely proportional to the square of mass resolution
(see equation (\ref{eq:den_res})). Similarly, the density resolution
limits in B/8 and Br/8 are 64 times smaller than B and Br.
The PDF histograms basically agree
very well below $n_{\rm H,res}$ as expected. 

In the runs  B/8 and Br/8, the density resolution limits are so low
that the high density regions ($\ga 100{\rm cm^{-2}}$) are not captured
properly. As a result, it seems to be difficult to distinguish the PDFs for
models B/8 and Br/8 from each other.
On the other hand, the results from the runs B/2 and Br/2 
indicate the same conclusion as we found in the runs B and Br,
since the density resolution is enough to capture the self-shielded dense
regions.
Thus, the disk in B/2 fragments due
to gravitational instability, whereas that in Br/2 is stable. Therefore we
conclude that the resolution of the present regular simulations 
are enough to capture 
the fragmentation of the disks under the assumptions we employed,
whereas in lower resolution runs such as B/8 and Br/8 are not able to
describe the present physical situation.

\section{Discussion}
\label{discussion}
It should be emphasized that present results directly prove the
suppression of star formation activities by ultraviolet background radiation
{\it in halos with $T_{\rm vir} > 10^4$K.} It has already been
known that the star formation is suppressed in less massive halos with
$T_{\rm vir} \la 10^4$K, since the gas in the dark halo potential
evaporates due to the high thermal pressure of photoheated gas
\citep{UI84,Efs92,BR92,TW96,BL99,FT00,Gne00c} if the self-shielding effect is not important \citep{Kita00,Kita01,SU04a,SU04b}.
On the other hand, in the present simulations, the gas in the dark halo
do not evaporate because of the deep gravitational
potential. We find that even in such halos, star formation could be heavily
suppressed in case gas is configured to form disks with low column density.

Another important result found in the present calculation is the
presence of a clear boundary in the column density of the disk, below which
star formation is heavily  suppressed. 
According to the numerical results, the critical column density is
$N_{\rm H}\sim 1-4 \times 10^{21}{\rm cm^{-2}}$. This threshold is very interesting
mainly for two reasons. 
First, although the present simulations are performed with fixed dark
halo mass, we can try to convert the critical column density into dark halo
mass. 
The mass of the uniform disk with given $N_{\rm H}$ and disk radius $R_{\rm disk}$ is 
\begin{equation}
M_{\rm disk} = 6\times 10^9M_\odot
\left(\frac{R_{\rm disk}}{10{\rm kpc}}\right)^2
\left(\frac{N_{\rm H}}{2.5\times10^{21}{\rm cm^{-2}}}\right).
\end{equation}
Therefore, 
if the mass of the host dark halo is 7($\simeq\Omega_{\rm B}/\Omega$) times the disk mass,
the critical dark halo mass is $\sim 4 \times 10^{10}M_\odot$. 
This value is still smaller than the critical scale found by \citet{downsizing}
by a few factor, but 
if we take into account the internal feedback effects such as UV
radiation from internal sources (AGN, massive stars) or supernova, the
threshold might account for the observed critical down-sizing mass.
We also point out that if we start the simulation from cosmological
setup, the threshold mass could be raised more, since photoheating might
be able to penetrate deeper into the disk because of initially less
dense configurations.
Thus, inclusion of such physics as well as starting simulations from
cosmological setup are necessary to obtain more precise understanding of
star formation in forming disk galaxies.

Secondly, the critical column density is as large as the star formation
threshold found in local disk galaxies \citep{K89,MK01}. In fact,
the observed star formation threshold of galactic disks has been
investigated by \citet{Schaye01,Schaye04} from theoretical side. 
He suggest that the galactic disk
could be gravitationally stable if it is unshielded from the external
UV radiation, since Q value of the photoheated disk
easily exceeds unity. 
As a result, gravitational fragmentation of
the disk is suppressed, so the star formation activities do.
\citet{SD08} performed simple hydrodynamics simulations taking into
account the star formation threshold column density using effective
equation of state in multiphase medium,
although they do not solve radiation transfer equations explicitly. They
found the threshold column density is as large as 
$4M_\odot {\rm pc}^{-2}$ (Fig.4 and 5 in \citet{SD08}), which is 
$5\times 10^{20}{\rm cm^{-2}}$ in cgs unit. This value is smaller
than that obtained in this paper by a few factor($\sim 2.5 \times 10^{21}{\rm
cm^{2}}$).
The basic mechanism they propose to suppress the star formation of the
disk is same as the one found in the present simulations, except that we do
not include the effects of dust extinction required
especially for present-day disk galaxies.
The gas disk is more
easily self-shielded by dust absorption, since the dust opacity at the
Lyman limit frequency is as large as HI continuum for solar metallicity.
Thus, it is it is reasonable that we have larger critical column
density than obtained in \citet{SD08}.
On the other hand, in the present simulations, we assume $Z=10^{-2}Z_\odot$.
Therefore, the dust opacity at Lyman limit is smaller than HI continumm
opacity by two orders of magnitude. As a result, dust opacity has much
smaller impact on the self-shielding effects at such low metallicity.
In any case, we would dare to mention that present
results almost succeeded to probe the presence of star formation threshold column density
proposed by \citet{Schaye04} utilizing the full 3D radiation
hydrodynamics simulations. More realistic calculations for present-day 
galaxies including the effects of dust extinction are left for future
works.

It is also worth noting that star formation rate (SFR) of the disk above the
threshold column density is consistent with the observed value, although 
it is difficult to compare the present results directly with observed
SFR, since we do not take into account the local
stellar feedbacks, the effects of dust particles, metal enrichment
followed by the radiative cooling by abundant metals.
Despite of such issues, 
SFR cloud be obtained by the density PDF assuming the stars are formed in
dense clumps within local free-fall time. It is given as
\begin{equation}
\dot{\Sigma_*}=\epsilon_{\rm c}\sqrt{G\rho_{\rm c}}\Sigma_{\rm disk}f_{\rm c}
\end{equation}
where $\epsilon_{\rm c}$ is the star formation efficiency, $\rho_{\rm
c}$ is the threshold density above which the gas is converted to
stars, $\Sigma_{\rm disk}$ denotes the surface density of the disk, and
$f_{\rm c}$ denotes the mass fraction of gas in the disk condensed into
dense clumps with $\rho > \rho_{\rm c}$.  $f_{\rm c}$ is the quantity
which can be obtained by integrating the density PDF above $\rho_{\rm
c}$. In the present set of simulations with radiative feedback, model Cr
is the only one in which star formation is expected since self-shielded
cold fragments emerges in the run. If we use $\epsilon=0.1$ and $n_{\rm
H,res}$ to assess the threshold density $\rho_{\rm c}$, 
we obtain $f_{\rm c}\simeq 0.066$. Using the surface density of the disk
in model Cr, the SFR is evaluated as  
$\dot{\Sigma_*}=0.03 {M_\odot {\rm yr}^{-1} {\rm kpc}^{-2}}$.
Although we cannot discuss the dependence of SFR on $\Sigma_{\rm disk}$,
the value is consistent with the observation \citep{K98}. This
reasonable agreement infer the validity of the present numerical models.

We also point out another standpoint on this issue by
\citet{SU00a},  in which they suggested that
self-shielding from ultraviolet background 
could be a key mechanism to determine the morphology of
galaxies\citep{SU00b}, although their arguments based upon 1-dimensional
radiation hydrodynamics calculations. Unfortunately, it is impossible to
relate the present results to morphology bifurcation of galaxies,  
since we assume a disk by hand in our simulations performed so far. From
this point of view, again we need to perform simulations from cosmological
initial conditions.

 

\section{Conclusion}
\label{conclusion}
In this paper, we perform radiation hydrodynamics simulations on the
fragmentation of galactic disks under the ultraviolet
radiation background. 
We find that ultraviolet radiation field strongly suppress
the star formation in the disks in case the photoheating is not shielded
enough. We emphasize that this suppression is different from
photoevaporation effect, 
because the rotation velocities at the outer boundary of the disks 
in the present set of
simulations are $\sim 100$km/s, 
which are fast enough to retain the photoheated gas.
In our simulations, we find a threshold column density of the
disk ($\sim 10^{21}{\rm cm^{-2}}$) above which the fragmentation is not suppressed. It is similar to the star formation threshold
column density observed in nearby galaxies.  
Presence of such critical threshold would
be one of the reason for the so-called down-sizing problem in nearby
galaxies.

\bigskip
HS thanks the anonymous referee and 
K. Wada for helpful comments and discussions.
HS also thanks
M. Umemura, M. Ohta and H. Sato for their continuous encouragements. 
The analysis has been made with computational facilities 
at Center for Computational Science in University of Tsukuba,  Konan
University, and Rikkyo University. 
This work was supported in part by Inamori Research Foundation as well
as Ministry of Education, Culture,
Sports, Science, and Technology (MEXT), Grants-in-Aid, Specially
Promoted Research 16002003.


\begin{table}[ht]
\begin{center}
\caption{Common Parameters\label{tab:common}}
\begin{tabular}{ccc}
\tableline\tableline
  symbol & numerical value employed & description  \\
\tableline
 $N_{\rm nei}$ & 50 & Number of neighbor particles \\
 $H_{\rm i}$ & 100pc & Initial disk thickness \\
 $R_{\rm disk}$ & 3kpc & Initial disk radius \\
 $T_{\rm i}$ & $10^4$K & Initial disk temperature \\
 $T_{\rm min}$ & $300$K & minimum temperature  \\
\tableline
\end{tabular}
\end{center}
\end{table}

\begin{table}[ht]
\begin{center}
\caption{Model dependent Parameters \label{tab:models}}
\begin{tabular}{ccccccc}
\tableline\tableline
Label& $I_{21}$ & $\rho_{\rm i}[M_\odot/{\rm pc^{-3}}]$ & simulated time
 & \# of SPH particles &$n_{\rm H,res}[{\rm cm^{-3}}]$&$\epsilon[{\rm pc}]$\\
\tableline
 A & 0 & 0.05 & 300 Myr & $1.28\times 10^6$ & 235 & 3.05\\
 B & 0 & 0.1 & 120 Myr & $2.56\times 10^6$ & 235 & 3.05\\
 C & 0 & 0.3 & 40 Myr & $7.68\times 10^6$ & 235 & 3.05\\
 Ar & 1 & 0.05 & 350 Myr & $1.28\times 10^6$ & 235 & 3.05\\
 Br & 1 & 0.1 & 200 Myr & $2.56\times 10^6$ & 235 & 3.05\\
 Cr & 1 & 0.3 & 40 Myr & $7.68\times 10^6$ & 235 & 3.05\\
\tableline
 B/2 & 0 & 0.1 & 120 Myr & $1.28\times 10^6$ &58.8& 6.09\\
 Br/2 & 1 & 0.1 & 120 Myr & $1.28\times 10^6$ &58.8& 6.09\\
\tableline
 B/8 & 0 & 0.1 & 120 Myr & $3.20\times 10^5$ &3.67& 24.4\\
 Br/8 & 1 & 0.1 & 120 Myr & $3.20\times 10^5$ &3.67& 24.4\\
\tableline
\end{tabular}
\end{center}
\end{table}


\setcounter{figure}{0}

\clearpage
\begin{figure}[ht]
\begin{center}
\includegraphics[angle=0,width=8cm]{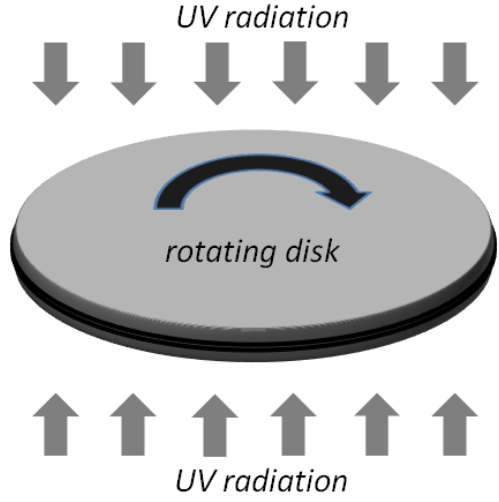}
\caption{Schematic view of the assumed configuration. Rotating gas disk is
 embedded in a dark halo potential, which is irradiated by the
 ultraviolet background radiation. The light rays are assumed to be
 perpendicular to the disk.
}\label{fig:model}
\end{center}
\end{figure}
\begin{figure}[ht]
\begin{center}
\includegraphics[angle=0,width=8cm]{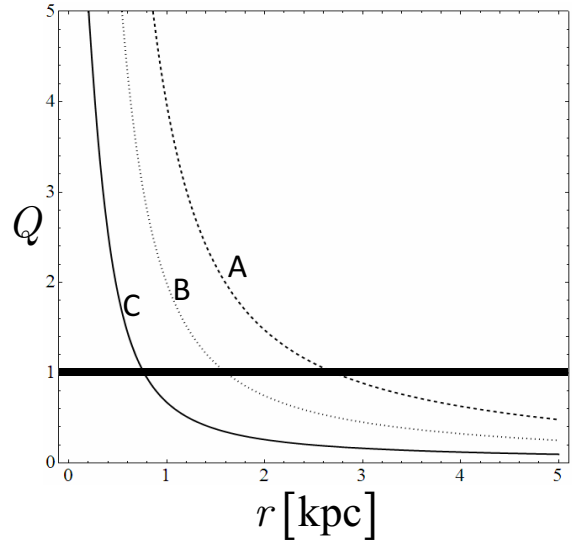}
\caption{
$Q(=c_{\rm s}\kappa/\pi G \Sigma)$ value is plotted as functions of radial
 coordinate in the disk. In case $Q > 1$ is satisfied in a certain region
 of a disk, the region is gravitationally stable, otherwise not.Three
 curves correspond to the three models A,B and C assuming $T=T_{\rm min}$.
}\label{fig:Q}
\end{center}
\end{figure}

\begin{figure}[ht]
\begin{center}
\includegraphics[angle=0,width=15cm]{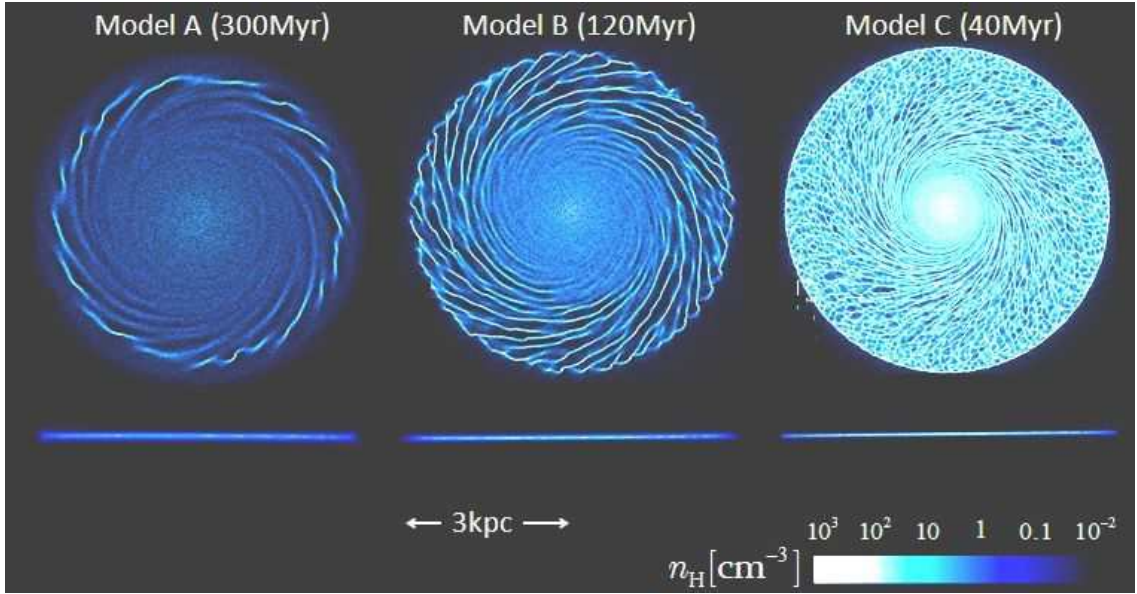}
\caption{Face-on/edge-on view of the disks in three models A B
 and C are shown. These are the snapshots taken at $t=300$Myr for model
 A, $t=120$Myr for model B and $t=40$Myr for model C, respectively. The
 color represents the density contrast, as shown in the legend.
}\label{fig:montage_norad}
\end{center}
\end{figure}

\begin{figure}[ht]
\begin{center}
\includegraphics[angle=0,width=15cm]{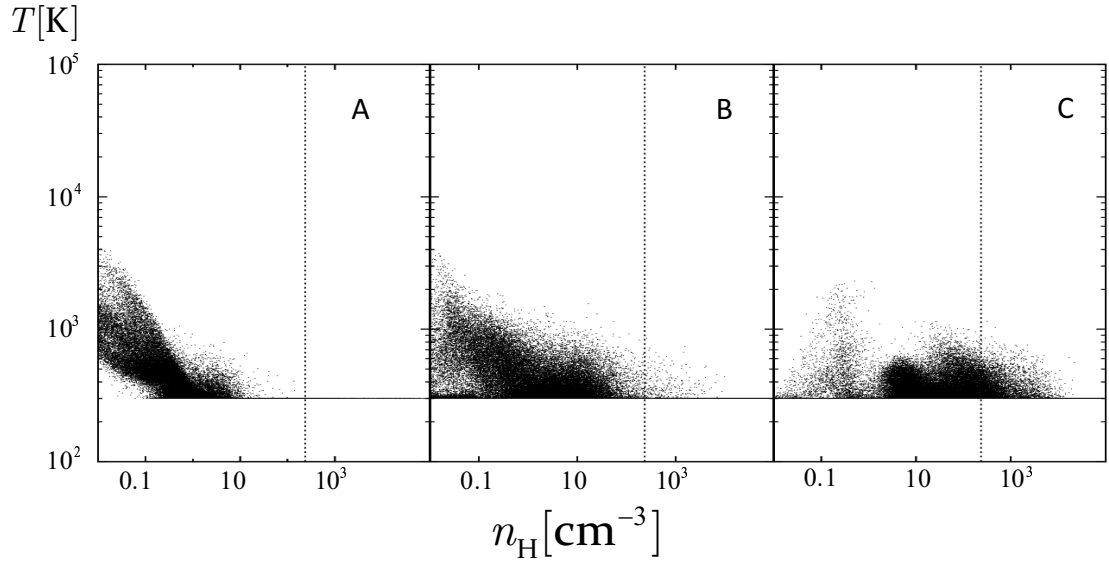}
\caption{Phase diagram for models A,B and C. Each dot represents each
 SPH particle. The vertical dotted lines represent $n_{\rm H,res}$. 
}\label{fig:phase_norad}
\end{center}
\end{figure}

\begin{figure}[ht]
\begin{center}
\includegraphics[angle=0,width=15cm]{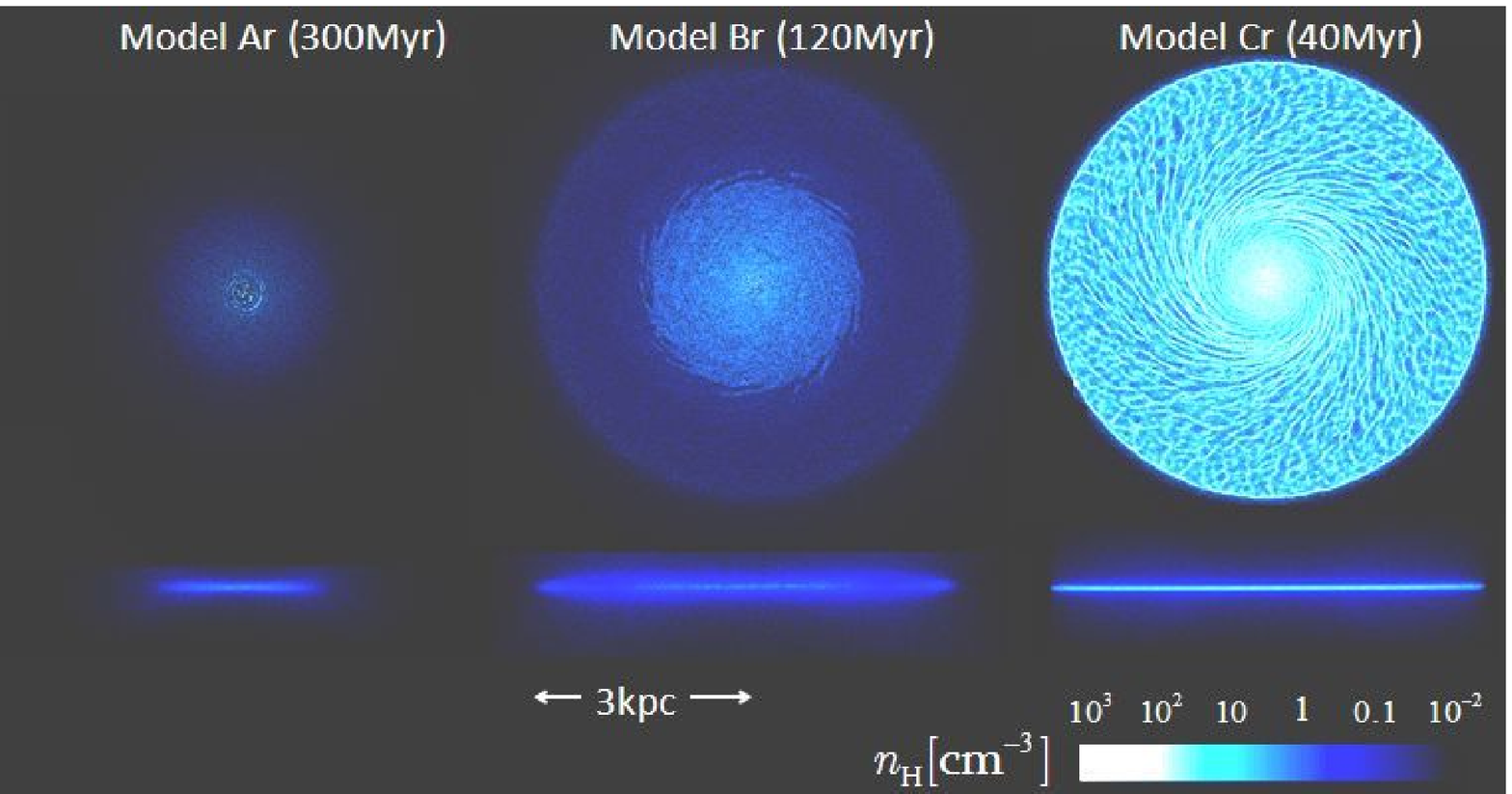}
\caption{Same as Fig.\ref{fig:montage_norad}, except that the models are
 Ar, Br and Cr.
}\label{fig:montage_rad}
\end{center}
\end{figure}
\begin{figure}[ht]
\begin{center}
\includegraphics[angle=0,width=15cm]{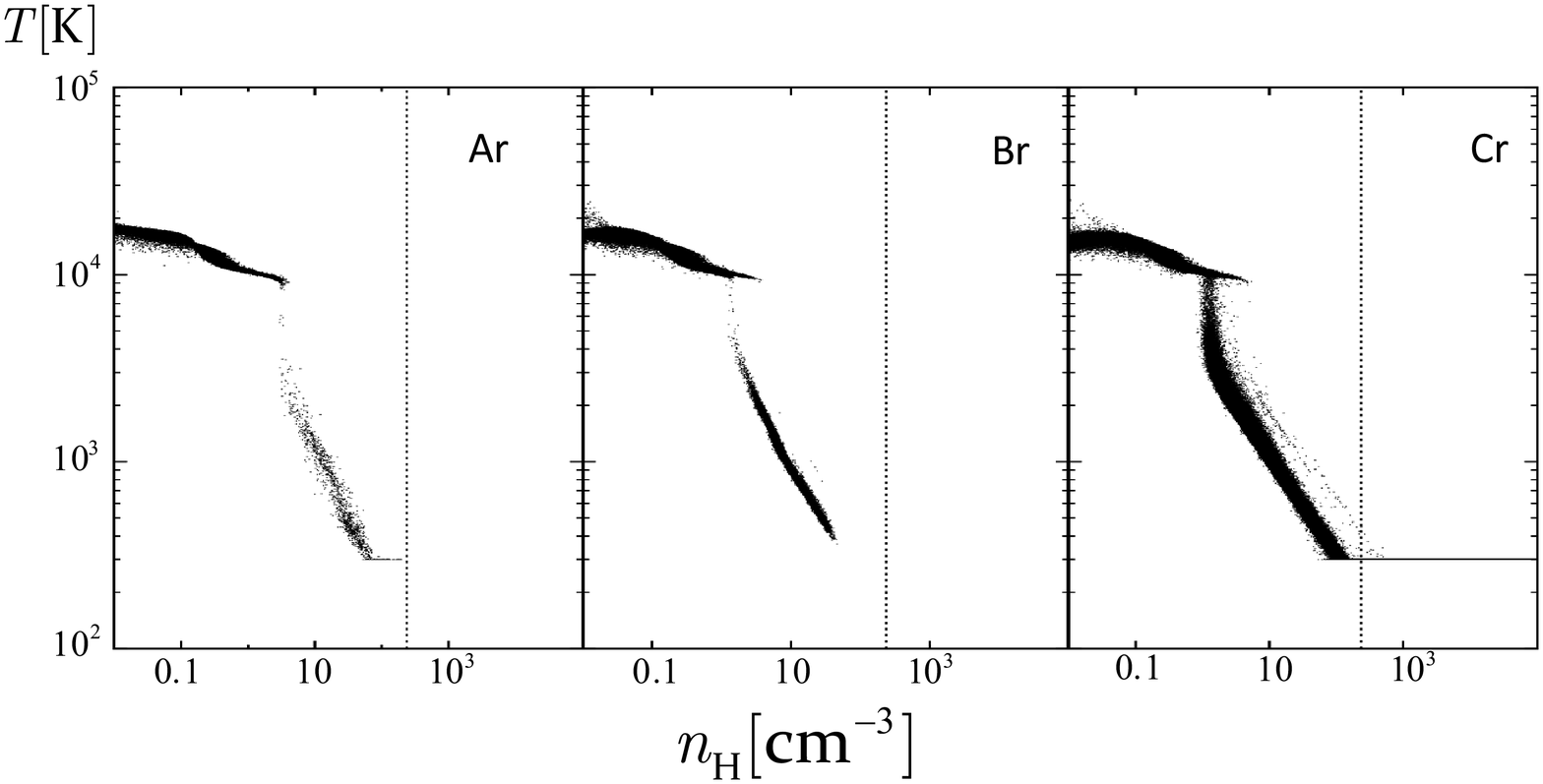}
\caption{Same as Fig.\ref{fig:phase_norad}, except that models are Ar,
 Br and Cr.
}\label{fig:phase_rad}
\end{center}
\end{figure}
\begin{figure}[ht]
\begin{center}
\includegraphics[angle=0,width=13cm]{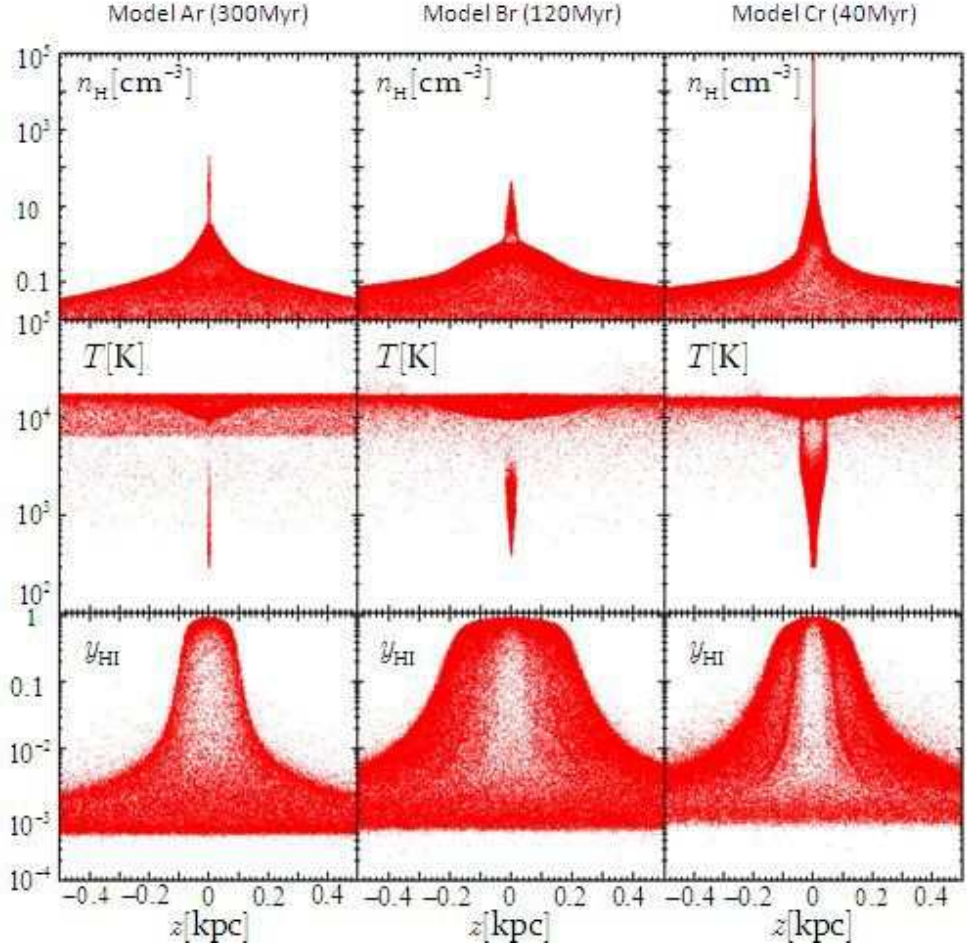}
\caption{Various quantities in the disks are plotted as functions of
 vertical coordinate, $z$. The horizontal axes denote $z$, whereas
 vertical axes denote density (top), temperature (middle) and HI
 fraction $y_{\rm HI}$ (bottom). These are the snapshots taken at
 $t=300$Myr for model A, $t=120$Myr for model B and $t=40$Myr for model
 C, respectively. 
}\label{fig:z-X}
\end{center}
\end{figure}
\begin{figure}[ht]
\begin{center}
\includegraphics[angle=0,width=13cm]{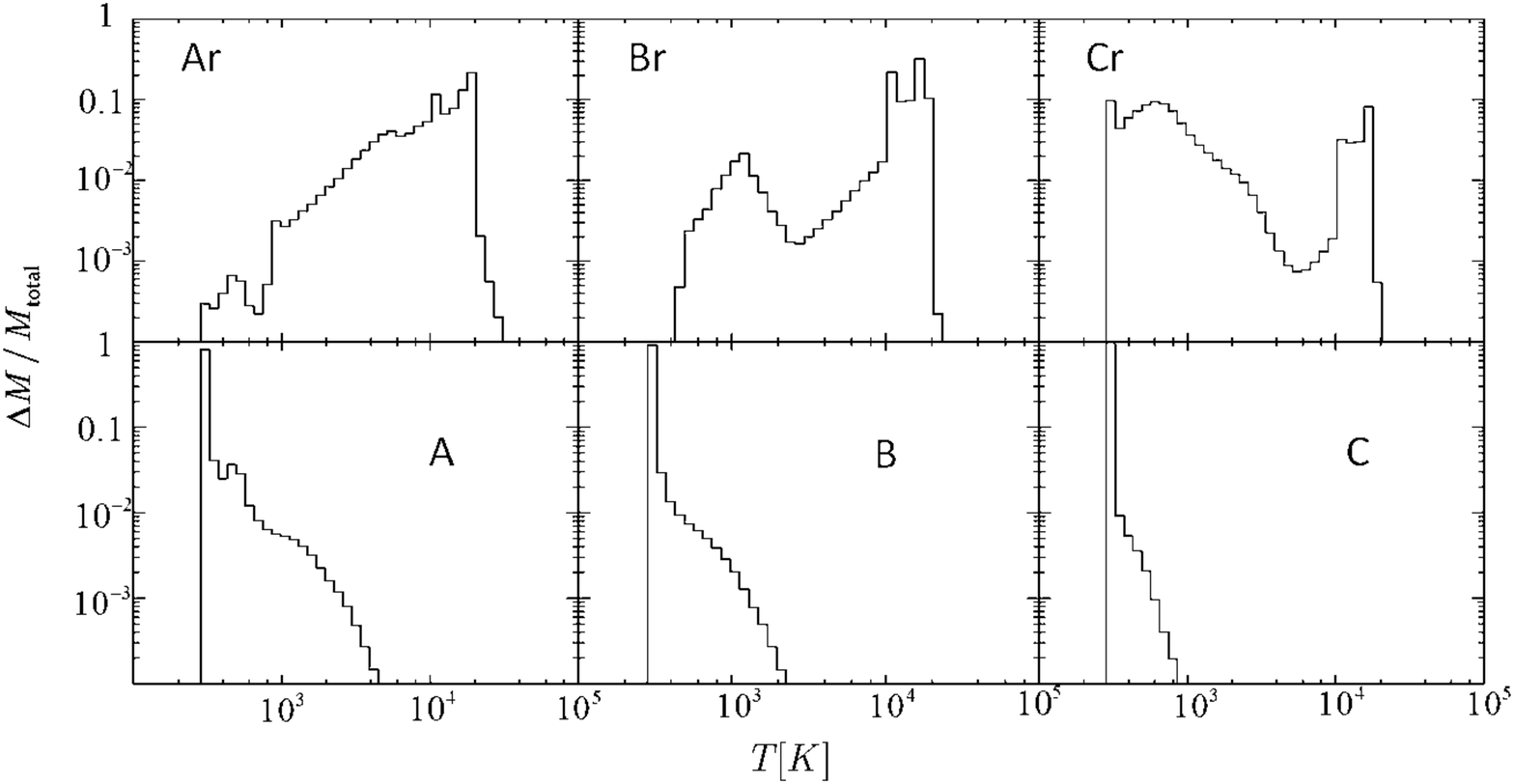}
\caption{Temperature Probability Distribution Functions(PDFs) are shown
 for six models Ar,Br,Cr,A,B and C. 
Horizontal axes denote the temperature, whereas the
 vertical axes show the differential mass fraction $\Delta M/M_{\rm total}$
 found within a logarithmic temperature bin $\Delta(\log T)=0.06$. 
Here $M_{\rm total}$ is the total gas mass of the
 disk. The histograms denote temperature PDFs at $t=300$Myr for models
 Ar and A, $t=120$Myr for models Br and B, $t=40$Myr for models Cr and C,
 respectively. 
}\label{fig:pdf_tmp}
\end{center}
\end{figure}
\begin{figure}[ht]
\begin{center}
\includegraphics[angle=0,width=15cm]{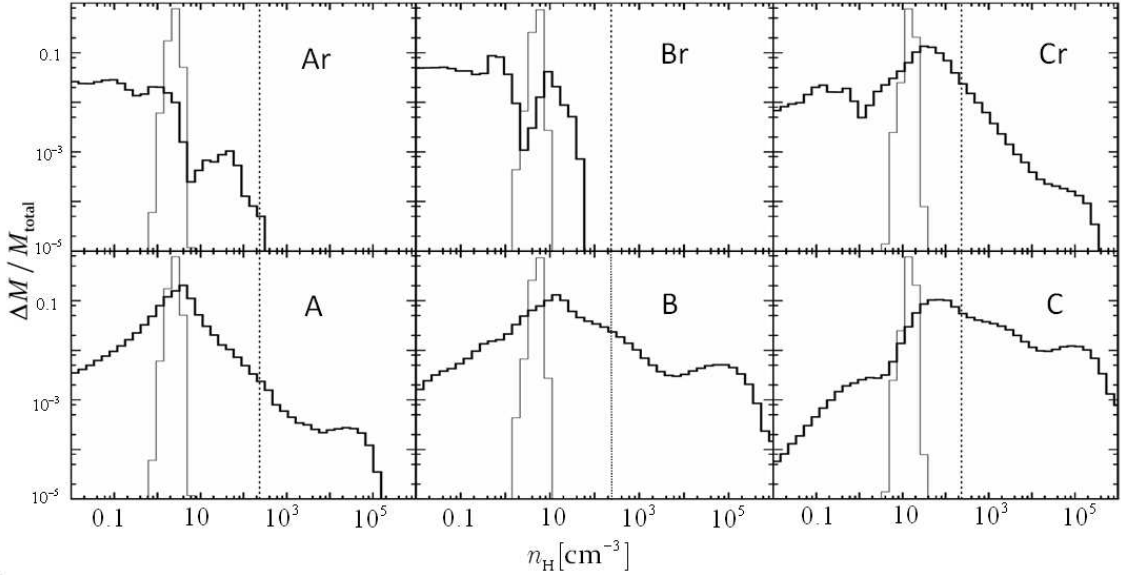}
\caption{Density Probability distribution functions(PDFs) are shown for models A,B,C,Ar,Br and Cr. Horizontal axes denote the density, whereas the
 vertical axes show the differential mass fraction $\Delta M/M_{\rm total}$
 found within a logarithmic density bin $\Delta(\log n_{\rm H})=0.18$. 
 The histogram drawn by thin solid lines denote the initial PDFs in
 the models. The thick solid lines denote the evolved PDFs at $t=300$Myr for models A and
 Ar, $t=120$Myr for models B and Br, $t=40$Myr for models C and Cr,
 respectively. The vertical dotted thin lines show the resolution limits of
 the runs.
}\label{fig:pdf}
\end{center}
\end{figure}
\begin{figure}[ht]
\begin{center}
\includegraphics[angle=0,width=15cm]{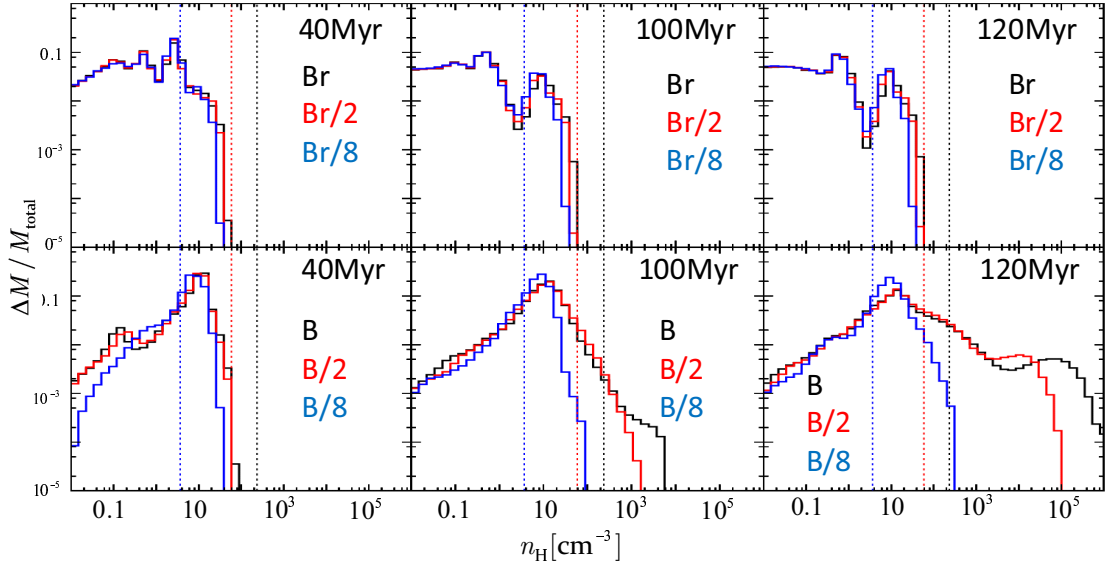}
\caption{PDFs in models B and Br are compared to the results from the
 low resolution runs(B/2,Br/2,B/8,Br/8). The axes are the same as
 Fig.\ref{fig:pdf}. Top three panels :the PDFs for models Br(black histogram)
 , Br/2(red), Br/8(blue) at $t=40$Myr(left), $100$Myr(middle) and $120$Myr(right).Density resolution limits, $n_{\rm H, res}$, corresponding to the runs
 Br, Br/2 and Br/8 are shown as the vertical dotted lines with colors
 same as the histgrams.
 Bottom panels : PDFs for models B, B/2 and B/8 at three epochs. The
 colors are same as the top panels.
}\label{fig:conv}
\end{center}
\end{figure}


\begin{thebibliography}{}
\bibitem[Altay et al.(2008)]{Altay08} Altay, G., Croft, 
R.~A.~C., \& Pelupessy, I.\ 2008, ArXiv e-prints, 802, arXiv:0802.3698 
\bibitem[Barkana \& Loeb(1999)]{BL99} 
Barkana, R., \& Loeb, A.\ 1999, \apj, 523, 54 
\bibitem[Bate \& Burkert(1997)]{BB97} 
Bate, M.~R., \& Burkert, A.\ 1997, \mnras, 288, 1060 
\bibitem[Boehringer \& Hensler(1989)]{BH89} 
Boehringer, H., \& Hensler, G.\ 1989, \aap, 215, 147 
\bibitem[Babul \& Rees(1992)]{BR92} Babul, A.~\& Rees, M.~J.\ 1992, \mnras, 255, 346 
\bibitem[Baldry et al.(2006)]{Baldry06} 
Baldry, I.~K., Balogh, M.~L., Bower, R.~G., Glazebrook, K., Nichol, R.~C., Bamford, S.~P., 
\& Budavari, T.\ 2006, \mnras, 373, 469 
\bibitem[Corbelli et al.(1997)]{Corbelli} Corbelli, E., Galli, 
D., \& Palla, F.\ 1997, \apjl, 487, L53 
\bibitem[Cowie et al.(1996)]{Cowie96} 
Cowie, L.~L., Songaila, A., Hu, E.~M., \& Cohen, J.~G.\ 1996, \aj, 112, 839 
\bibitem[Dalgarno \& McCray (1972)]{DM73} Dalgarno, A. \& McCray, A. 1972, ARA\&A, 10, 375 
\bibitem[Draine \& Bertoldi (1996)]{DB96}
Draine, B. T., \& Bertoldi, F. 1996, \apj, 468, 269
\bibitem[Efstathiou(1992)]{Efs92} Efstathiou, G.\ 1992, \mnras, 256, 43P
\bibitem[Erb et al.(2006)]{Erb06} 
Erb, D.~K., Steidel, C.~C., Shapley, A.~E., Pettini, M., Reddy, N.~A., 
\& Adelberger, K.~L.\ 2006, \apj, 647, 128 
\bibitem[Ferrara \& Tolstoy(2000)]{FT00} Ferrara, A.~\& Tolstoy, E.\ 2000, \mnras, 313, 291 
\bibitem[Ferland(2000)]{CLOUDY} Ferland, G.~J.\ 2000, Revista 
Mexicana de Astronomia y Astrofisica Conference Series, 9, 153 
\bibitem[Galli \& Palla (1998)]{GP98} 
Galli D. \& Palla F. 1998, \aap, 335, 403 
\bibitem[Gnedin (2000)]{Gne00c} Gnedin, N.~Y.\ 2000, \apj, 542, 535 
\bibitem[Gnedin \& Abel (2001)]{GA01}
Gnedin, N. Y.\& Abel, T , 2001, NewA, 6, 437
\bibitem[Iliev et al.(2006)]{TSU3} Iliev, I.~T., et al.\ 
2006, \mnras 371, 1057
\bibitem[Kauffmann et al.(2003)]{downsizing} Kauffmann, G., et 
al.\ 2003, \mnras, 341, 54 
\bibitem[Kennicutt(1989)]{K89} Kennicutt, R.~C., Jr.\ 1989, \apj, 344, 685 
\bibitem[Kennicutt(1998)]{K98} Kennicutt, R.~C., Jr.\ 1998, 
\apj, 498, 541 
\bibitem[Kitayama et al.(2000)]{Kita00} Kitayama, T., Tajiri, 
Y., Umemura, M., Susa, H., \& Ikeuchi, S.\ 2000, \mnras, 315, L1 
\bibitem[Kitayama et al.(2001)]{Kita01} Kitayama,T., Susa, H., Umemura, M., \& Ikeuchi, S. 2001, \mnras, 326, 1353
\bibitem[Kodama et al.(2004)]{downsizing_subaru} Kodama, T., et al.\ 
2004, \mnras, 350, 1005 
\bibitem[Martin \& Kennicutt(2001)]{MK01} Martin, C.~L., \& Kennicutt, R.~C., Jr.\ 2001, \apj, 555, 301 
\bibitem[Mellema et al.(2006)]{C2ray} Mellema, G., Iliev, 
I.~T., Alvarez, M.~A., \& Shapiro, P.~R.\ 2006, New Astronomy, 11, 374 
\bibitem[Omukai(2000)]{Omukai00} Omukai, K.\ 2000, \apj, 534, 
809 
\bibitem[Papovich et al.(2006)]{Papovich06} 
Papovich, C., et al.\ 2006, \apj, 640, 92 
\bibitem[Qiu et al.(2007)]{Qiu07} 
Qiu, J.-M., Feng, L.-L., Shu, C.-W., \& Fang, L.-Z.\ 
2007, New Astronomy, 12, 398 
\bibitem[Reddy et al.(2006)]{Reddy06} 
Reddy, N.~A., Steidel, C.~C., Fadda, D., Yan, L., Pettini, M., Shapley, A.~E., Erb, D.~K., \& Adelberger, K.~L.\ 2006, \apj, 644, 792 
\bibitem[Rijkhorst et al.(2006)]{Saru} Rijkhorst, E.-J., Plewa, T., Dubey, A., \& Mellema, G.\ 2006, \aap, 452, 907 
\bibitem[Robertson \& Kravtsov(2007)]{Robertson07} Robertson, B., \& Kravtsov, A.\ 2007, ArXiv e-prints, 710, arXiv:0710.2102 
\bibitem[Saitoh et al.(2008)]{Saitoh08} Saitoh, T.~R., Daisaka, 
H., Kokubo, E., Makino, J., Okamoto, T., Tomisaka, K., Wada, K., 
\& Yoshida, N.\ 2008, ArXiv e-prints, 802, arXiv:0802.0961 
\bibitem[Schaye(2001)]{Schaye01} Schaye, J.\ 2001, \apjl, 562, L95 
\bibitem[Schaye(2004)]{Schaye04} Schaye, J.\ 2004, \apj, 609, 667 
\bibitem[Schaye \& Dalla Vecchia(2008)]{SD08} Schaye, J., \& Dalla Vecchia, C.\ 2008, \mnras, 383, 1210 
\bibitem[Shapiro \& Kang (1987)]{SK87}
Shapiro, P.R., \& Kang, H., 1987, \apj, 318, 32
\bibitem[Spitzer(1978)]{Spitzer78}
Spitzer, L. Jr. 1978, in Physical Processes in the Interstellar Medium
(John Wiley \& Sons, Inc. 1978)
\bibitem[Songaila (2001)]{Son01} Songaila, A. 2001, \apj, 561, 153L
\bibitem[Steinmetz \& M\"uller(1993)]{SM93} Steinmetz, M.~\&  M\"uller, E.\ 1993, \aap, 268, 391 
\bibitem[Susa \& Kitayama (2000)]{SuKi00} 
Susa, H. \& Kitayama, T. 2000, \mnras, 317, 175
\bibitem[Susa \& Umemura(2000a)]{SU00a} Susa, H., \& Umemura, M.\ 2000, \apj, 537, 578 
\bibitem[Susa \& Umemura(2000b)]{SU00b} Susa, H., \& Umemura, M.\ 2000, \mnras, 316, L17 
\bibitem[Susa \& Umemura(2004a)]{SU04a} 
Susa, H. \& Umemura, M. 2004, \apj, 600, 1
\bibitem[Susa \& Umemura(2004b)]{SU04b} 
Susa, H. \& Umemura, M. 2004, \apj, 610, 5L
\bibitem[Susa (2006)]{Susa06} 
Susa, H. 2006, \pasj, 455, 58
\bibitem[Susa \& Umemura(2006)]{SU06} 
Susa, H. \& Umemura, M. 2006, \apj, 645, 93L
\bibitem[Susa(2007)]{Susa07} Susa, H.\ 2007, \apj, 659, 908 
\bibitem[Tasker \& Bryan(2008)]{Tasker08} Tasker, E.~J., \& Bryan, G.~L.\ 2008, \apj, 673, 810 
\bibitem[Thacker et al. (2000)]{Thac00} 
Thacker, R.J., Tittley, E.R., Pearce, F.R., Couchman, H.M.P. \& Thomas, P.A. 2000, 
\mnras 319, 619
\bibitem[Thoul \& Weinberg(1996)]{TW96} Thoul, A.~A.~\& 
Weinberg, D.~H.\ 1996, \apj, 465, 608 
\bibitem[Toomre(1964)]{Toomre} Toomre, A.\ 1964, \apj, 139, 1217 
\bibitem[Umemura \& Ikeuchi(1984)]{UI84} Umemura, M.~\& Ikeuchi, S.\ 1984, Progress of Theoretical Physics, 72, 47 
\bibitem[Umemura(1993)]{Ume93} Umemura, M.\ 1993, \apj, 406, 361 
\bibitem[Wada \& Norman(2007)]{Wada07} Wada, K., \& Norman, C.~A.\ 2007, \apj, 660, 276 
\bibitem[Whalen \& Norman(2008)]{Whalen08} Whalen, D.~J., \& Norman, M.~L.\ 2008, \apj, 672, 287 
\bibitem[Young \& Lo (1997a)]{YL97a} Young, M. \& Lo, Y. 1997, \apj, 476, 127
\bibitem[Young \& Lo (1997b)]{YL97b} Young, M. \& Lo, Y. 1997, \apj, 490, 710
\bibitem[Yoshida et al.(2007)]{Yoshida07} Yoshida, N., Oh, S.~P., 
Kitayama, T., \& Hernquist, L.\ 2007, \apj, 663, 687 
\end{thebibliography}
\end{document}